\documentclass[12pt,preprint]{emulateapj}

\newcommand{\kms}{$\rm km s^{-1}$} 
\newcommand{\ho}{$\rm H_o = 73 km s^{-1} Mpc^{-1}$} 
\newcommand{\dm}{$\Delta m_{15}(B)$} 
\newcommand{\dmgz}{$\Delta m_{15}(B) = 0.69 \pm 0.04$} 

\begin{document}

\shortauthors{Hicken et al.}
\slugcomment{Accepted for publication in ApJL}
\shorttitle{The Luminous and Carbon-Rich SN~2006gz}

\title{The Luminous and Carbon-Rich Supernova 2006gz:  A Double Degenerate Merger?}

\author{M. Hicken\altaffilmark{1}, 
P. M. Garnavich\altaffilmark{2},
J. L. Prieto\altaffilmark{3},
S. Blondin\altaffilmark{1},
D. L. DePoy\altaffilmark{3}, 
R. P. Kirshner\altaffilmark{1}, 
J. Parrent\altaffilmark{4}
}

\altaffiltext{1}{Harvard-Smithsonian Center for
Astrophysics,Cambridge, MA 02138; mhicken, sblondin,
kirshner@cfa.harvard.edu}  \altaffiltext{2}{Dept. of Physics,
University of Notre Dame, Notre Dame, IN 46556; pgarnavi@nd.edu}
\altaffiltext{3}{Dept. of Astronomy, The Ohio State University,
Columbus, OH 43210; prieto, depoy@astronomy.ohio-state.edu}
\altaffiltext{4}{Dept. of Physics and Astronomy, University of
Oklahoma, Norman, OK 73019; parrent@nhn.ou.edu}

\begin{abstract}

Spectra and light curves of SN 2006gz show the strongest signature of
unburned carbon and one of 
the slowest fading light curves ever seen in a type~Ia event (\dmgz).  
The early-time \ion{Si}{2} velocity is low, implying  
it was slowed by an envelope of unburned material.
Our best estimate of the luminosity implies $M_V = -19.74$ 
and the production of $\sim1.2 M_{\sun}$ of $^{56}$Ni.  This suggests
a super-Chandrasekhar mass progenitor.  A double degenerate merger 
is consistent with these observations.

\end{abstract}


\section{Introduction}

The conventional picture for Type~Ia supernovae (SN Ia) is that they
are thermonuclear explosions of C-O white dwarf stars (WD).  Two 
possible routes to explosion have been explored:  single degenerate (SD)
models in which the WD is nudged to explosion by accretion from a 
binary companion, and double degenerate (DD) models in which two WDs
merge.  

SN Ia have been the essential tool in establishing the Hubble constant
\citep[e.g.,][]{jha99,riess05}, showing a sub-critical matter
density \citep{garnavich98,perlmutter98}, and demonstrating that the
cosmic expansion is accelerating
\citep[e.g.,][]{riess98,perlmutter99}. Yet the precise nature
of SN Ia progenitors remains uncertain.

In the SD scenario, there is only a narrow range of 
physical conditions for accretion to increase the WD mass over time.
\citep[see][and references therein]{yoon03}.  The DD pathway may 
lead to neutron star formation \citep{saio04}, 
but recent studies suggest explosion as
a SN Ia is possible \citep{yoon07}.  Both scenarios may
contribute to the SN Ia population so it is important to consider 
their observable differences.

The presence or absence of carbon and its velocity distribution
should be strong clues to the nature of the explosion.  
For example, multi-dimensional calculations of pure deflagrations 
predict that unburned material should be mixed into 
low-velocity layers \citep[e.g.,][]{gamezo03}. The fact that carbon is
not commonly seen in deep layers suggests that most SN Ia are 
not a result of pure deflagration.

Carbon is most often weak or absent in optical and infrared spectra of
SN Ia \citep{branch03, marion06, thomas06}.  This suggests  that
burning to intermediate mass elements takes place in the 
outer layers of a C-O WD for
most SN Ia. The model that matches this feature best is a  delayed
detonation in a SD system since DD merger and pulsating delayed
detonation (PDD) explosions are likley to leave  significant unburned
material in the outer layers \citep*{khokhlov93}.  Conversely, the
presence of carbon raises the possibility of a DD merger.  DD mergers
could also have masses and luminosities that extend beyond the
Chandrasekhar limit.  Two recent SN Ia may have been merger events:
the extremely  luminous SN~2003fg was interpreted by \citet{howell06}
as arising  from a super-Chandrasekhar (S-Ch) progenitor, and a
compressed silicon layer in SN~2005hj provides indirect evidence 
for an unburned layer atop the silicon \citep{quimby07}.

Here we present spectra and light curves\footnote{Available at
http://www.cfa.harvard.edu/supernova/SN2006gz} of the type~Ia
supernova 2006gz that begin two weeks before maximum light.  They
directly show
the strongest  signature of unburned carbon ever reported and a low
silicon  velocity at very early times. The light curve is  very broad
and luminous, approaching the luminosity of SN~2003fg.  These properties
suggest that SN~2006gz  may have arisen from the merger of two WDs.

\section{Observations}

SN~2006gz was discovered by \citet{puckett06} and independently by
\citet{winslow06} on 2007 Sept. 26.0 (UT). It was classified as a
type~Ia event by \citet{prieto06a} who noted an unusual double
absorption feature instead of the typical Si II 6355\AA~line. The
supernova was 12\arcsec ~west and 28\arcsec ~south of the center of
the Scd spiral galaxy IC~1277, corresponding to a projected
distance of 14.4 kpc.  It should be noted that the 21-cm \ion{H}{1}
radio-based  redshift of 0.0280 \citep{haynes97} is incorrect.  We
report z=0.0237  $\pm$0.0004 based on our optical spectrum of
IC~1277, in close  agreement with other optical redshifts reported in
NED\footnote{http://nedwww.ipac.caltech.edu/}.

Seven spectra were obtained with the 2.4-m MDM telescope and The
Ohio State University Boller and Chivens CCD spectrograph beginning
2006 Sept.~28.1 (UT), 14 days before maximum light.  A 2\arcsec~slit 
and a 150 line/mm grating were used.  The resolution was $\sim$15\AA.

A series of 10 spectra was taken with the Fred L. Whipple Observatory 
(FLWO) 1.5-m Tillinghast telescope and FAST spectrograph \citep{fabricant98}, 
also starting Sept. 28.1 (UT).  A 3\arcsec~slit 
and 300 line/mm grating were mounted.  The resolution was $\sim$7\AA.

Beginning Sept. 29.1 (UT), 23 nights of $UBVr'i'$ 
photometry were aquired on the FLWO 1.2m telescope and Keplercam 
instrument.
Eleven comparison stars were calibrated on 4 photometric nights in $BV$, 
3 nights in $r'i'$, and 1 night in $U$, with a typical $V$-band 
uncertainty of 0.014 mag per star.  Host galaxy subtraction images 
were taken in $BVr'i'$ on 2007 May 27.4 (UT) and in $U$ on June 27.4 
(UT), after SN~2006gz had faded sufficiently.

\section{Analysis}

\subsection{Spectra}

\begin{figure}
\plotone{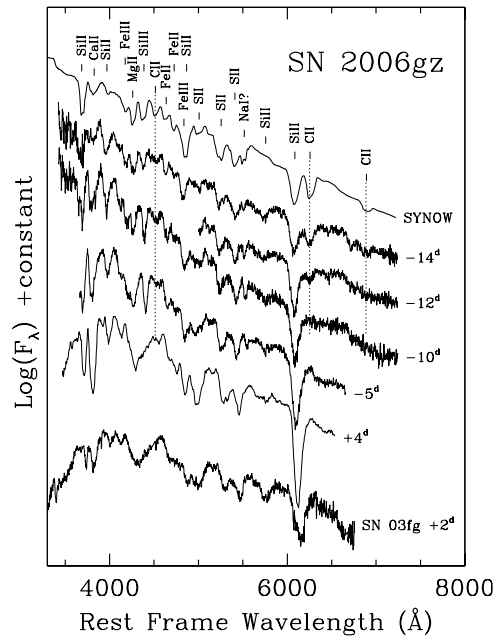}
\caption{SN~2006gz spectra at 5 epochs.  SN~2003fg at +2 days 
and a SYNOW fit to the MDM spectrum at -14 days
are shown.  Dotted lines:  C II $\lambda\lambda$4745, 6580, and 7234
blueshifted by 15500 \kms. }
\label{fig_spec}
\end{figure}

\begin{figure}
\plotone{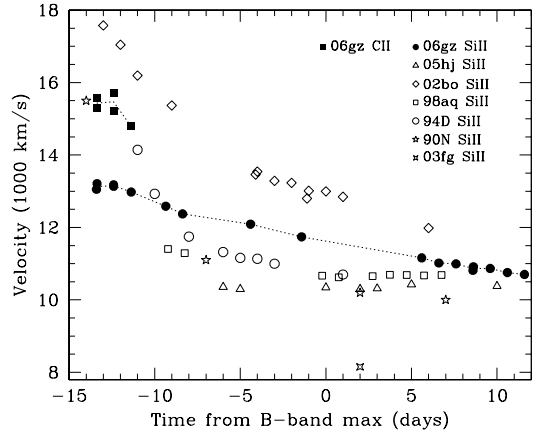}
\caption{\ion{Si}{2} and \ion{C}{2} velocities.  Up until 
--12 days, \ion{Si}{2} is suppressed in SN~2006gz while its 
\ion{C}{2} velocity is comparable to that of \ion{Si}{2} 
in SN~1994D and 1990N.  SN~2006gz velocities are consistent 
within the error bars (not shown) for nights with measurements
from both telescopes.}
\label{fig_vel}
\end{figure}

The FLWO and MDM spectra taken at --14 days
both show features typical of a type~Ia supernova. As seen in 
Figure~\ref{fig_spec}, intermediate-mass elements (\ion{Si}{2}, 
\ion{S}{2} and \ion{Ca}{2}) are clearly present.
\citet{prieto06a} noted that the most unusual aspect of the early spectra is the
extra absorption seen redward of the \ion{Si}{2} 6355\AA~feature.
Absorption features at this wavelength have been identified
with \ion{C}{2} 6580\AA~\citep{branch03,thomas06}.  Using the SYNOW code
to generate synthetic spectra \citep{fisher99}, \ion{C}{2} and \ion{Na}{1}
reproduce this feature and the absorption at 
5490\AA~better than a combination of \ion{H}{1} 
and \ion{He}{1}.  We adopt this interpretation.
\ion{C}{2} lines at 4745\AA~and 7234\AA~may also 
be seen before -10 days.  At two weeks before
maximum light the \ion{C}{2} feature has a rest-frame equivalent width (EW) of
25\AA~in SN~2006gz, while in SN~1990N \citep{leib91,jeffery92} the feature 
is seen at 5\AA~EW. \citet{thomas06} reports a strong \ion{C}{2} detection 
in SN~2006D with 7\AA~EW at --7 days.
In the case of SN~2006gz, the \ion{C}{2} absorption strength is comparable
to the \ion{Si}{2} EW of 37\AA~at --14 days but quickly weakens and is
undetectable past --10 days (Figure~\ref{fig_vel}).
The \ion{Si}{2} EW increases after the initial spectrum and reaches 70\AA~near
maximum brightness.

Unlike the luminous and slowly declining SN 1991T, the spectra before
maximum light are not dominated by Fe III lines. But the ratio of
\ion{Si}{2} 5972 to the 6355\AA~line seen here does correspond to the
values seen in other very slowly declining SN Ia \citep[e.g.,][]{garnavich04}.
Except for the 5972\AA~feature, many of the \ion{Si}{2} lines 
in SN~2006gz are particularly
deep and narrow, including 3858\AA~and 4130\AA. There is also a strong \ion{Si}{3} line
(4560\AA) and \ion{Si}{3} may also be the source of the absorption observed at 5500\AA~\citep{branch03}.

We measure the \ion{Si}{2} velocity to be $\sim$13100 \kms~at --14 and --13 days, in contrast to $\sim$15500 \kms~in
SN~1990N at --14 days (Figure~\ref{fig_vel}).  But the SN~2006gz \ion{C}{2}
velocity is similar to the silicon velocity in 
SN~1990N and SN~1994D before --11 days \citep{patat96}.  The \ion{Si}{2} 
velocity does not show the rapid decline that was observed in other SN Ia
before --7 days.  At --10 days it is at $\sim$12500 \kms~and declines by
only 1000 \kms~to 11500 \kms~by maximum light.  In four cases of SN Ia 
caught in early phases (1990N, 1994D, 2002bo, and 2003du) the \ion{Si}{2} 
velocities at $t<-10$ days were 2000 to 4000 \kms~higher than that 
at maximum light \citep{benetti05}. 
In SN~2006gz, the velocity of the \ion{Si}{2} is unusually low and constant
at the earliest times, possibly as a result of overlying material for which
the \ion{C}{2} line is evidence.  
A tantalizing hint of an early plateau in velocity is suggested by the
unchanging velocities in both \ion{Si}{2} and \ion{C}{2} between days --14
and --13.  After --7 days, the \ion{Si}{2} velocity slowly declines,
with a slope similar to SN~1990N and SN~1994D.

In the DD merger models (and PDD models)
studied by \citet{khokhlov93}, the detonation of the dense C-O occurs
within a low-density C-O envelope. The unburned envelope compresses
the outer, partially burned material, creating a density enhancement in
the silicon layer which slows the photosphere from receding into
lower-velocity layers.  In SN~2005hj, \citet{quimby07} inferred 
the presence of overlying material from the
observed velocity plateau, but they did not have direct evidence for a
carbon-rich layer.  In SN 2006gz, carbon is definitely present, but
the possible velocity plateau is very brief and very early.

\subsection{Light Curve, Reddening and Luminosity}

\begin{figure}
\plotone{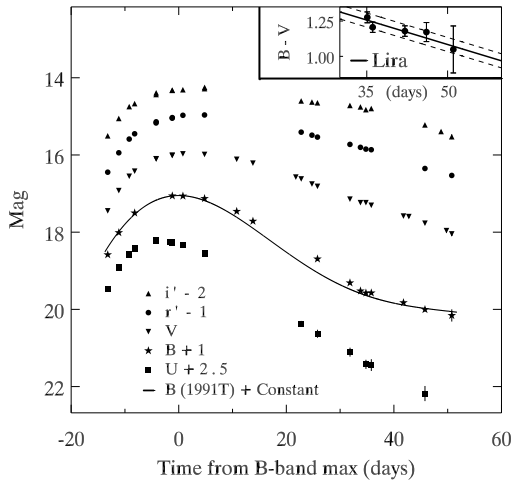}
\caption{SN~2006gz $UBVr'i'$ light curves.  $B$-band has a faster rise
and slower decline than SN~1991T.  Inset:  $B-V$
from +35 to +51 days, with $E(B-V)_{Gal} = 0.063$ removed.  
The Lira relation, offset by 
$E(B-V)_{host} = 0.18 \pm 0.05$, is shown.}
\label{fig_lc}
\end{figure}

SN~2006gz has one of the broadest light curves (Figure~\ref{fig_lc})
of any SN Ia ever measured:  its 15-day $B$-band decline from maximum
light is \dmgz, with $t_{max}(B) = 2454020.2 \pm 0.5$ JD.  The rise 
time of 16.6 $\pm$ 0.6 days is very short for the slow decline and
does not fit the positive correlation between rise and decline time in 
light curve fitters such as MLCS2k2 \citep{jha07} and dm15 \citep{prieto06b}.  
\citet{strovink07} find evidence for a bimodal SN Ia rise time distribution
based on a sample of 8 objects:  $16.64 \pm 0.21$ days and $18.81 \pm 0.36$ 
days.  The decline rate is also outside the range used to construct these 
fitters.  The  $i'$-band peaks several days after $B$ and does not appear to 
have the pronounced trough or second peak expected for a very luminous SN Ia.
For comparison, SN~2001ay has \dm $\sim 0.6-0.7$ but its luminosity of 
$M_V \sim -19.2$ is significantly lower than expected 
for its decline rate\citep{phillips02}.
Slow-declining and overluminous SN Ia are found almost exclusively in late-type
galaxies while fast-declining and subluminous SN Ia are usually found in  
early-type hosts \citep[e.g.,][]{jha07}.  \citet{mannucci06} show that there
may be 2 populations of SN Ia progenitors:  a promptly-exploding component
that dominates the supernova rate in star-forming galaxies,  
and a "tardy" one that gives rise to most of the SN Ia in older hosts.
How the SD and DD pathways may relate to the correlation between luminosity
and host type, and to the prompt and tardy progenitor components, needs
further investigation.

The luminosity and extinction of SN~2006gz in $UBV$ are given in Table 
~\ref{table_mag}.  We correct for Galactic reddening using the dust maps of
\citet{schlegel98} and apply $K$-corrections.  We use a Hubble-flow distance 
modulus of $\mu = 34.95 \pm 0.04$ ($z_{cmb} = 0.0234$, \ho, $\Omega_M=0.3$, 
$\Omega_\Lambda=0.7$, and a peculiar velocity uncertainty of 300 \kms) 
to get absolute luminosities of $M_B=-19.17 \pm 
0.04$, and $M_V=-19.19 \pm 0.04$, before correction for host reddening.

There are, however, 2 pieces of evidence for host
reddening.  
First, the \ion{Na}{1}
absorption at the SN position has an $EW=0.30 \pm 0.15 $\AA, 
giving rise to $E(B-V)_{host} \le 0.15 \pm 0.08$ \citep{turatto02}.
Second, the $B-V$ color shows normal, 
albeit slow, evolution.  It increases by $\sim$ 1.3 mag between maximum
light and 35 days after.
We apply $K$-corrections and correct for time dilation and Galactic
reddening in  the $B-V$ color curve. From +35 to +51 days, SN 2006gz
has a color  evolution consistent with the Lira relation
\citep{lira95}, implying a host color excess of $E(B-V)_{host}=0.18
\pm 0.05$ (Figure~\ref{fig_lc}).   We adopt this while cautioning that
it may be incorrect if Lira's relation does not apply.  Since the spectra 
are normal after --10 days, and the Lira-corrected  $B-V$ and $U-B$ 
colors at maximum are consistent with those reported  in Figure 9 of
\citet{jha06}, it is not unreasonable to assume that Lira's relation
applies.  

The correct value of the reddening law for SN Ia is under
active study 
\citep{conley07}.  We use $R_V = 3.1$ 
and $R_V = 2.1$ to give a wide range of plausible 
host extinctions and we calculate
absolute magnitudes for SN~2006gz (see Table~\ref{table_mag}).  Using
$R_V = 3.1$ gives $M_B = -19.91 \pm 0.21$, and $M_V=-19.74\pm0.16$,
with $B-V=-0.17 \pm 0.05$.  This is comparable to the possibly S-Ch
SN~2003fg, which had an absolute magnitude of $M_V=-19.85 \pm 0.06$
(\ho)  and a color of  $B-V \sim -0.15$, implying its host reddening
was  quite small.  The case of $R_V=2.1$ gives $M_V=-19.56 \pm 0.11$
for SN~2006gz.

\begin{figure}
\plotone{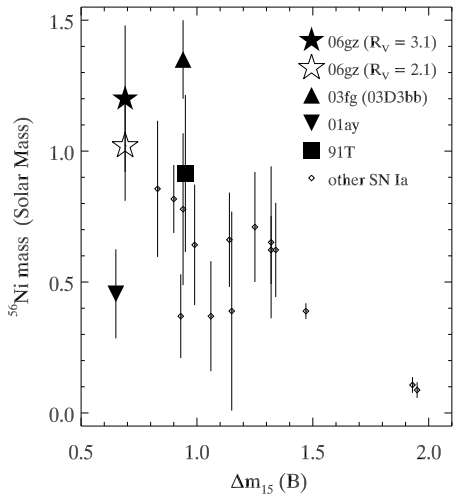}
\caption{$^{56}$Ni mass v. \dm for both reddening cases of SN~2006gz.  
Also shown are 16 SN Ia from \citet{stritz06},
and SN~2003fg \citep{howell06}, and SN~2001ay \citep{phillips06}.  
SN~2006gz and SN~2003fg may come from S-Ch progenitors.}
\label{fig_ni}
\end{figure}

We also calculated a UVOIR light curve of SN~2006gz with $\rm L_{bol}(t_{max}) =(2.18 \pm
0.39)~10^{43}~erg~s^{-1}$ for $R_V=3.1$.  We follow the
procedure of \citet{stritz06} to derive a $^{56}$Ni  mass ($M_{\rm Ni}$) of $1.20 \pm 0.28M_{\sun}$.  
For $R_V = 2.1$,  $M_{\rm Ni} = 1.02 \pm 0.21M_{\sun}$. 
The detonation of a $1.4M_{\sun}$ WD can produce 
$\sim0.9-1.0M_{\sun}$ of  $^{56}$Ni (Khokhlov, A. 2007, private 
communication).
In figure~\ref{fig_ni}, we plot $M_{\rm Ni}$  against
\dm for the 16 SN Ia in \citet{stritz06} and add SN~2006gz,
SN~2003fg, and SN~2001ay \citep{phillips06}.

The DD merger models of \citet{khokhlov93} consist of a 1.2
$M_{\sun}$ WD detonation inside C-O envelopes of different masses.
They produce $0.63~M_{\sun}$ of $^{56}$Ni.   The envelope decelerates
the layers of the exploded WD, increases their density,  and thus
raises the diffusion time.  This produces broader light curves, as
seen in SN~2006gz after maximum.  Merger models with larger, S-Ch 
primary WDs should be explored to see
if they can explain the properties of SN~2006gz.  A different
cause of the fast rise time and high luminosity could be an
off-centered explosion, with the nickel-rich region towards us
\citep{hillebr07}.

\section{Discussion and Conclusion}

The three outstanding features of SN~2006gz are the presence of
carbon, the low early-time silicon velocity, and the luminous, broad light
curve.

SN~2006gz shows the strongest evidence for unburned carbon of any
SN Ia observed to date. The strong carbon feature has a higher
expansion velocity than the silicon and fades quickly.  The carbon
is part of the photosphere at early times.  The obvious inference is that 
the explosion did not burn all the carbon.  This
could be explained by the outer WD C-O layer being accelerated 
and diluted during a deflagration phase.  It is also 
possible that the envelope of the DD merger could be
shocked and accelerated within the first few seconds of explosion
and give rise to the observed carbon.

The unusually low and slowly declining silicon velocity at early 
times, combined with the deep and narrow silicon absorption 
features, have been predicted by DD merger models.  

The light curve is one of the broadest SN Ia ever seen with \dmgz.  If it
obeys the Lira relation then it produced $M_{\rm Ni}=1.20 \pm0.28 M_{\sun}$~($R_V=3.1$)
or $M_{\rm Ni}=1.02 \pm0.21 M_{\sun}$~($R_V=2.1$).
Combined with this, the presence of copious silicon and 
other non-iron-peak elements 
rules out a SD detonation and makes a S-Ch progenitor a good 
possibility.  In both SD and DD cases, the centrifugal force of rotation
can allow WDs to grow beyond $1.4 M_{\sun}$.  However, the DD scenario
is simpler in the sense that no helium or hydrogen shell burning is 
required to build up the primary WD.

Taken separately, the three main features of SN~2006gz can be explained
by various models.  First, the unburned carbon is 
plausible to different degrees in the SD and DD scenarios.  Second, 
the suppressed silicon velocity is predicted by DD models.  Third, 
the luminous and broad 
light curve with high implied $^{56}$Ni mass is possible in SD and DD models.
However, taking all of these features together points to a consistent 
picture in which SN~2006gz is the result of the merger of two white dwarfs.

\acknowledgments

We thank M. Calkins for obtaining the very early carbon-rich FLWO 1.5m spectra  
and T. Currie and W. Peters for additional 1.5m spectra.
Thanks to L. Watson and J. Eastman for MDM 2.4m spectra, and to P. Berlind,
G. Esquerdo, M. Everett, J. Fernandez, I. Ginsburg, D. Latham, and 
A. Vaz for FLWO 1.2m photometry.  We are grateful to R. Quimby and R. C. 
Smith for providing digital spectra of published supernovae.  We thank 
P. H\"oflich for many discussions, and the Aspen Center for Physics and the 
KITP at the University of California, Santa Barbara.  This work has been 
supported, in part, by NSF grants AST0606772 to Harvard University and PHY0551164.


\begin{deluxetable}{ccccc|cc|cc}
\tablecolumns{9}
\tablewidth{0pc}
\tablecaption{Extinction and Absolute Magnitude at Maximum Light with $E(B-V)_{host}$ = 0.18(05)}
\tablehead{
\colhead{} & \colhead{} & \colhead{} & \colhead{} & \colhead{$A_X$(host)=0} &
\multicolumn{2}{c}{$R_V=2.1$} & \multicolumn{2}{c}{$R_V=3.1$} \\
\colhead{X-band} & \colhead{m$_X$} & \colhead{$A_X(Gal)$} & \colhead{$K$-cor} &
\colhead{M$_X$} & \colhead{$A_X$(host)} & \colhead{M$_X$}  & 
\colhead{$A_X$(host)} & \colhead{M$_X$}
}
\startdata
U & 15.77 & 0.34 & 0.17 & -19.70 & 0.67(19) 
& -20.37(19) & 0.87(24) & -20.57(25) \\
B & 16.06 & 0.27 & 0.01 & -19.17 & 0.56(16) 
& -19.73(16) & 0.74(21) & -19.91(21) \\
V & 15.99 & 0.21 & 0.02 & -19.19 & 0.38(11) 
& -19.56(11) & 0.56(16) & -19.74(16) \\
\enddata
\label{table_mag}
\end{deluxetable}

\clearpage

\end{document}